\documentclass[aps,prb,twocolumn,groupedaddress,nofootinbib]{revtex4-2}
\usepackage[utf8]{inputenc}
\usepackage{amsmath}
\usepackage{amsfonts}
\usepackage{amssymb}

\usepackage{graphicx}
\usepackage[left=2cm,right=2cm,top=2cm,bottom=2cm]{geometry}

\usepackage[version=3]{mhchem}
\usepackage{siunitx}
\DeclareSIUnit\bar{bar}
\DeclareSIUnit\angstrom{\text {Å}}
\DeclareSIUnit\barn{\text {barn}}

\usepackage{url}

\usepackage[labelformat=parens,caption=false]{subfig}

\usepackage[hidelinks]{hyperref}
\usepackage[capitalize]{cleveref}

\usepackage{todonotes}

\newcommand{\abinitio}{\emph{ab initio}}
\newcommand{\betagox}{$\beta$-{\ce{Ga2O3}}}
\newcommand{\alphasic}{$\alpha$-\ce{SiC}}
\newcommand{\sicTwoH}{\ce{SiC}-2H}
\newcommand{\sicFourH}{\ce{SiC}-4H}
\newcommand{\sicSixH}{\ce{SiC}-6H}

\newcommand{\codename}[1]{{\sc {#1}}}
\newcommand{\abins}{\codename{abins}}
\newcommand{\castep}{\codename{castep}}
\newcommand{\euphonic}{\codename{euphonic}}
\newcommand{\ibex}{\codename{ibex}}
\newcommand{\mantid}{\codename{mantid}}
\newcommand{\oclimax}{\codename{oclimax}}
\newcommand{\phonopy}{\codename{phonopy}}
\newcommand{\scipy}{\codename{scipy}}

\newcommand{\kpoint}{\(\mathbf{k}\)-point}

\newcommand{\qpoint}{\(\mathbf{q}\)-point}
\newcommand{\qpoints}{\qpoint{}s}

\newcommand{\suppl}{Supplementary Material}

\usepackage[acronym]{glossaries}
\setacronymstyle{long-short}
\glsdisablehyper  
\newacronym{ccr}{CCR}{closed-cycle refrigerator}
\newacronym{dft}{DFT}{density-functional theory}
\newacronym{dfpt}{DFPT}{density-functional perturbation theory}
\newacronym{dos}{DOS}{density of states}
\newacronym{gga}{GGA}{generalised gradient approximation}
\newacronym{ir}{IR}{infra-red}
\newacronym{ins}{INS}{Inelastic neutron scattering}
\newacronym{lda}{LDA}{local-density approximation}
\newacronym{metagga}{meta-GGA}{meta-GGA}
\newacronym[longplural={machine-learned interatomic potentials},shortplural={MLIPs}]{mlip}{MLIP}{machine-learned interatomic potential}
\newacronym[longplural={norm-conserving pseudopotentials},shortplural={NCPs}]{ncp}{NCP}{norm-conserving pseudopotential}
\newacronym{tof}{ToF}{time-of-flight}
\newacronym{xc}{XC}{exchange--correlation}
\newacronym{xrd}{XRD}{x-ray diffraction}

\begin{document}

\title{Molecular neutron spectroscopy techniques applied to ceramics \alphasic{} and \betagox{}}
\author{Adam J. Jackson}
\email[Contact author: ]{adam.jackson@stfc.ac.uk}
\affiliation{Scientific Computing, UKRI STFC, Rutherford Appleton Laboratory, Didcot, OX11 0QX, UK}

\author{Svemir Rudi\'{c}}
\author{Manh Duc Le}
\author{Sanghamitra Mukhopadhyay}\email[Contact author: ]{sanghamitra.mukhopadhyay@stfc.ac.uk}
\affiliation{ISIS Neutron and Muon Source, UKRI STFC, Rutherford Appleton Laboratory, Didcot, OX11 0QX, UK}

\date{\today}

\begin{abstract}
    Neutron spectroscopy is a powerful technique for determining the
    vibrational states of matter.  In order to obtain a well-resolved spectrum
    in a reasonable amount of beam-time, instruments with fixed geometry
    may measure inelastic scattering at a limited set of angles,
    producing a 1--D spectrum $S(\omega)$.
    Such measurements are usually simulated in a DOS-like semi-analytic incoherent approximation,
    well-established for study of bending/stretching modes in molecular crystals. 
    In this work we empirically test the simulation method for two commercially-obtained ceramics with industrial electronic applications that act as ``worst-case'' systems.
    The phonon scattering from \alphasic{} and \betagox{} is coherent, depends on momentum transfer $Q$ and sits in frequencies below the typical ``fingerprint'' range of molecular spectroscopy.
    Inelastic neutron-scattering measurements of powders were performed with two contrasting spectrometers at cryogenic and elevated temperatures, and simulations performed using a variety of density-functional approximations.
    We find that for 1--D powder spectra from a compact instrument, the approximate simulations are easily comparable with experimental spectra 
    and give similar results to a more computationally-intensive numerical sampling of the coherent spectrum.
    Given the success with these systems,
    the approximate method appears to be suitable for modelling inelastic neutron scattering by harmonic phonons of almost any powder sample with this technique.
    When a $Q$-resolved instrument is used to collect the 2--D dynamical structure factor $S(Q,\omega)$, numerical averaging is still required to capture phonon features.
    Our simulations of inelastic scattering from \alphasic{} in the 6H polytype using the PBEsol functional gave good agreement with the experiments.
    By contrast, the RSCAN functional gave the best agreement with the measured spectra of \betagox{}
    and is recommended for future work on the lattice dynamics of this material.
    \vspace{1cm}
\end{abstract}

\maketitle

\section{Introduction}\label{sec:intro}
\gls{ins} is a vibrational spectroscopy technique, complementary to optical or x-ray measurements.
Due to the relatively large neutron mass, there can be significant momentum transfer probing phonon excitations away from the $\Gamma$-point of crystalline materials and inaccessible to \gls{ir} or Raman spectroscopy.
As neutrons interact directly with the nuclei of the vibrating atoms rather than their electronic shells (except for a magnetic interaction in cases of unfilled shells),
the technique is generally insensitive to electronic structure and instead depends on isotope-specific nuclear cross-sections.
With a total scattering cross-section of \SI{82.02}{\barn}, an order of magnitude greater than commonly-encountered elements, \ce{^1_1H+} nuclei tend to dominate the resulting spectra of hydrogenous materials.

Different classes of instruments and modelling assumptions are generally employed for molecular spectroscopy \emph{versus} the study of phonons in inorganic crystals.
In this work we measure and simulate the spectra of ceramic powders with both approaches,
in order to investigate the validity of the simulation approximations and validate \abinitio{} methods.

\subsection{Neutron spectrometers}
Inelastic neutron spectrometers use monochromating crystals, mechanical choppers or filters to choose or determine the neutron energy.
At continuous neutron sources such as nuclear reactors, generally the neutron energies are fixed before and after scattering from the sample in order to determine the sample energy transfer.
In contrast, if the neutron beam is pulsed the neutron \gls{tof} can be used to determine either the initial (before) or final (after scattering) neutron velocity assuming that all neutrons leave the source at one known instance.

Here we consider two types of instrument that have multiple implementations around the world:
an \emph{indirect-geometry} spectrometer using a crystal analyzer to fix the final neutron energy detected at a small set of scattering angles;
and a \emph{direct-geometry} spectrometer using a mechanical chopper to set the incident energy and detecting the scattering with a large bank of detectors.
In the former case, the relationship between \gls{tof} and energy is determined by the combined length between the source and analyzer;
this allows a compact instrument at a suitable distance from the pulsed source to have useful energy resolution and count rate.
This configuration is particularly suited to vibrational spectroscopy of hydrogenous samples and interest has grown in application to non-hydrogenous materials~\cite{parkerRecentFutureDevelopments2014a}.
The need for analyzer components between sample and detector makes it difficult to sample a wide range of scattering angles;
in turn this imposes strict ``kinematic constraints'' on the measured energy--momentum space.%
\footnote{An ingenious solution to the analyzer placement problem is embodied by the BIFROST instrument at the European Spallation Source~\cite{toft-petersenBIFROSTIndirectGeometry2025}.}
This makes it difficult to gather information about phonon dispersion, but is acceptable when measuring a 1--D $S(\omega)$ spectrum at large $Q$ values where powder-averaging gives a roughly uniform Brillouin Zone sampling.
(In this notation $Q$ is the scalar momentum \emph{transfer} in a scattering event whereas $\mathbf{q}$ is a specific vector in reciprocal space. $\omega$ is used for both phonon frequency and the equivalent measured energy transfer.)

Direct-geometry \gls{tof} spectrometers can include a large detector bank, gathering scattering information in a 4--D volume of $S(\mathbf{q}, \omega)$ space which is typically sliced or averaged for 2--D data presentation.
This is particularly relevant for the study of magnons and phonon dispersion in single crystals,
but the mechanical chopper design presents trade-offs in the achievable measurement resolution and count-rate.
Complementary measurements across both type of instrument can be useful for molecular spectroscopy, as in the high-$\omega$-low-$Q$ region direct-geometry chopper instruments can achieve superior resolution with reduced Debye--Waller intensity loss~\cite{ewingsUpgradeMAPSNeutron2019}.

\subsection{Established simulation methods}\label{sec:sim-methods-intro}
Planning and interpretation of these experiments is aided by atomistic simulations.
For the study of intramolecular vibrations in molecules and their crystals,
and of harmonic phonon spectra in crystalline solids,
harmonic force constants are calculated to useful accuracy from first principles using \gls{dft}.
Several computer codes are available for simulation of the corresponding INS spectrum (i.e. the dynamic structure factor $S(Q,\omega)$):
\abins{} supports a \gls{dos}-like incoherent approximation for powder-averaging including multi-phonon excitations~\cite{dymkowskiAbINSModernSoftware2018};
\euphonic{} can be used to calculate one-phonon coherent structure factor for a single crystal or with numerical powder-averaging~\cite{fairEuphonicInelasticNeutron2022a}
(there is also some experimental functionality in \phonopy{} for this case~\cite{togoFirstprinciplesPhononCalculations2023});
the \oclimax{} package supports both of these approaches~\cite{chengSimulationInelasticNeutron2019,chengCalculationThermalNeutron2020}.
For systems with modest anharmonicity, effective potential schemes and higher-order lattice dynamics can be used to account for some frequency renormalisation and lifetime broadening~\cite{hellmanLatticeDynamicsAnharmonic2011,knoopTDEPTemperatureDependent2024,erikssonHiphivePackageExtraction2019,kimNuclearQuantumEffect2018,heAnharmonicEigenvectorsAcoustic2020}.
Otherwise, to deal with severe anharmonicity and longer-timescale events it becomes more appropriate to use molecular dynamics and sample dynamical properties,
as implemented in codes such as \codename{nmoldyn}, \codename{mdanse}, \codename{dynasor} and \oclimax{}~\cite{hinsenNMoldyn3Using2012,goretMDANSEInteractiveAnalysis2017,franssonDynasorToolExtracting2021,bergerDynasor2Simulation2025,chengSimulationInelasticNeutron2020}.
In turn, experimental spectra can validate the atomistic modelling methods for further research:
in this work we use \gls{ins} measurements to make recommendations from a selection of \gls{xc} functionals.

Simulated \gls{ins} spectra for (generally hydrogenous) molecular spectroscopy
generally make a \gls{dos}-like incoherent ``almost-isotropic" approximation originating from the 1970s and 1980s.
Analytic powder averages were worked out for the power series of phonon order contributions to the incoherent dynamical structure factor;
the series is truncated to yield manageable leading terms with more complex mode-dependent Debye--Waller factors,
which may be simplified if the atomic displacements are approximately isotropic~\cite{thomasIncoherentInelasticNeutron1975,waddingtonInelasticNeutronScattering1982}.
As implemented in \abins{} and \oclimax{}, the higher-order terms are increasingly simplified with isotropic approximations;
this enables a computationally efficient treatment of the multi-phonon terms
which is not available to the coherent-scattering contribution.
In order to realise the benefits of this approach, one makes the ``incoherent approximation'' and assumes that the overall form of incoherent and coherent scattering are the same:
the incoherent contribution to $S(\mathbf{q}, \omega)$ is calculated for each atom and then weighted with a combined cross-section $\sigma_\mathrm{inc} + \sigma_\mathrm{coh}$.
It has been suggested that this is a good approximation at large $\mathbf{q}$ where there is little correlation~\cite{loveseyTheoryNeutronScattering1984}.
The ``\gls{dos}-like'' element of the approximation relates to the separation of the phonon wavevector \qpoints{} from the absolute value of momentum transfer $Q$ used in the INS intensity calculation;
intensities are computed for the modes based on their phonon frequencies/eigenvectors,
and independently the $Q$-dependence of the INS intensity is applied to the resulting spectrum.
While the method is applied to both classes of instruments considered here,
the neglect of phonon dispersion is quite noticeable in 2--D visualisations for large-area spectrometers
and more easily overlooked in 1--D spectra.

The impact of this stack of approximations has not been well examined for atypical samples on such instruments.
In this work we consider sample materials that this method was not intended for:
\ce{\alphasic} and \ce{\betagox},
as well as the \ce{Al} sample environment.
Direct-geometry measurements are simulated by numerical averaging from coherent scattering calculations;
for indirect-geometry, in addition to the established method described above, we examine the kinematically accessible part of the powder-averaged coherent \gls{ins} spectrum.
We also compare TOSCA measurements with a direct-geometry chopper spectrometer (MARI), which is more typically used for such systems.

\subsection{\alphasic{}}
\ce{SiC} is a wide-bandgap semiconductor adopting a range of stacking polytypes.
It has several desirable properties for power electronics applications: high thermal stability, resistance to large electric fields and operation at higher voltage than established \ce{Si}-based technology~\cite{iannacconePowerElectronicsBased2021}.
\ce{SiC} (primarily in the 4H polytype) is also being tested as a solid-state neutron-detecting material for nuclear reactors~\cite{coutinhoSiliconCarbideDiodes2021,ruddySiliconCarbideNeutron2022}.
The thermal conductivity, long established as ``rather high compared to other solids'', is strongly dependent on the sample purity and governed by phonon scattering~\cite{slackThermalConductivityPure1964}.
More recent electron-phonon coupling calculations (within the \gls{lda}) show that this contribution is not negligible at low temperature~\cite{protikPhononThermalTransport2017}.

Raman spectroscopy has been employed to study the vibrational spectrum of \ce{SiC} in various forms including
thin-film~\cite{hobertVibrationalSpectroscopySiC1999}, single-crystal (across multiple stacking polytypes)~\cite{feldmanPhononDispersionCurves1968}, and radiation-damaged crystals \cite{sorieul_raman_2006}.
The energy shifts of the phonon modes vary subtly between polytypes;
this has been exploited to construct joint dispersion curves from Raman spectroscopy of multiple polytypes~\cite{feldmanPhononDispersionCurves1968}.
The dispersion curves of \sicSixH{} have also been mapped with triple-axis neutron spectroscopy~\cite{dornerPhononDispersion6HSiC1998}.
This is a precise technique but rather slow; the \gls{ins} intensity is determined in the full 4--D $(\mathbf{q},\omega)$ space at chosen sampling points.

Two studies examine the sensitivity of SiC lattice parameters and bulk modulus to the choice of \gls{dft} \gls{xc} functional.
One of these is a survey over 44 solids and performs all calculations non-self-consistently on the converged PBE density~\cite{tranRungs142016}; the other study specifically considers SiC polytypes, their elastic parameters, and possible correction schemes~\cite{pizzagalliAccurateValues3C2021}.
In both cases PBEsol compares favourably with other GGAs while still giving slightly under-bound lattice parameters and bulk modulus, whereas LDA slightly overbinds (i.e. under-estimates the lattice constant)~\cite{tranRungs142016,pizzagalliAccurateValues3C2021}.
Meta-GGA functionals do not clearly deliver better results, and long-range dispersion corrections have a mixed impact.

In recent years researchers have trained \glspl{mlip} for SiC phases from DFT data, in order to explore size and time ranges beyond those feasible with \abinitio{} methods~\cite{liuDeepLearningInteratomic2024,zhangPhononThermalProperties2024,hamedaniSiCTGAPMachineLearning2025,chengRevealingPhononSignature2025}.
As well as the model architecture, hyperparameters and training structure diversity, the accuracy of such models is ultimately determined by the underlying reference: PBE appears to be the most popular \gls{xc} functional in this field but PBEsol has also been used.

\subsection{\betagox{}}
Gallium oxide has attracted recent interest as an ultra-wide bandgap semiconductor;
bulk single crystals can be produced in the thermodynamically-stable $\beta$ polymorph, and doped for \emph{n}-type conductivity~\cite{peartonReviewGa2O3Materials2018,zhangRecentProgressElectronic2020,higashiwakiVGa2O3MaterialProperties2022}.
The material is noted for a high thermal stability yet relatively low thermal conductivity in the range \SIrange{0.11}{0.27}{\watt\per\cm\per\kelvin} with strong anisotropy (and some measurement uncertainty); this is dominated by phonon transport at room temperature.
The thermal conductivity represents a challenge for dissipation in power electronics, but an opportunity for thermoelectric applications~\cite{guoAnisotropicThermalConductivity2015,handwergTemperaturedependentThermalConductivity2015}.
This contrasts with \alphasic{} which has high thermal stability \emph{and} high thermal conductivity;
\betagox{} has a more complex phonon band structure and broad phonon linewidths.
The $\Gamma$-point phonon modes of \betagox{} have been characterised by temperature-dependent polarized Raman spectroscopy~\cite{Dohy1982}.
Due to symmetry selection rules some of the modes are not visible in the Raman spectrum and were measured by IR spectroscopy over a smaller temperature range.
(In turn, the IR misses most of the Raman-visible modes.)
Force constants were fitted to data, allowing assignment of $\Gamma$-point modes to symmetry and movement.
In that work temperature-dependent frequency shifts were illustrated;
they appear modest up to room temperature but the Gr\"uneisen parameters were not calculated.
A variety of first-principles methods have been used for \betagox{} lattice dynamics:
in the harmonic approximation and up to third-order force constants using \gls{dft} with the PBE and PBEsol \gls{xc} functionals~\cite{liuLatticeDynamicalDielectric2007,jacksonOxidationGaNInitio2013,sunInfluencePolymorphismLattice2024};
with electron-phonon coupling in the \gls{lda}~\cite{ghoshInitioCalculationElectron2016};
and extended molecular dynamics simulations with a \gls{mlip} trained on PBE calculations~\cite{wangDissimilarThermalTransport2024}.

\section{Methods}
\subsection{X-ray diffraction}
Powdered \alphasic{} material was obtained commercially from Goodfellow: nominally this was \SI{99}{\percent} purity in ``hexagonal'' polytype.
A sample was characterized by \gls{xrd} using a Rigaku MiniFlex-600 system with a Cu-K$\alpha$ source and assigned to the 6H polytype.
The diffraction pattern was compared to simulations from the ICSD for reference structures in 2H, 4H and 6H polytypes,
with the 6H pattern giving the most convincing agreement~\cite{zagoracRecentDevelopmentsInorganic2019}.
(Diffraction patterns are plotted in \suppl{} \cite{supp:material}.)

\subsection{Inelastic neutron scattering}
Indirect-geometry \gls{ins} measurements were performed using the TOSCA neutron spectrometer at ISIS~\cite{colognesiTOSCANeutronSpectrometer2002}.
Samples of \alphasic{} (as above) and \betagox{} (99.99\%, Sigma Aldrich) were dried in a vacuum oven at \SI{70}{\celsius}.
Samples of \SI{9}{\gram} \alphasic{} and \SI{8}{\gram} \betagox{} were each loaded into an aluminium pouch and sample container on the TOSCA instrument.
In this design, samples are contained in a flat void of $48 \times 40 \times 2 \si{\milli\meter}$
to ensure effective usage of the beam cross-section while minimising the potential for multiple scattering.
The sample-holder sits in a \gls{ccr} operating below \SI{10}{\K}
 and the sample cell is electrically heated to regulate the sample temperature~\cite{downCryogenicSampleEnvironment2014}.
The system was purged with helium and a vacuum pump used to reduce the pressure to ${\sim\SI{30}{\milli\bar}}$.
Measurements were performed at the \gls{ccr} base temperature and with the cell heated to \SI{100}{\K} and \SI{200}{\K}, counting for $\sim\SI{15}{\hour}$ (\SI{2500}{\micro\ampere\hour} integrated proton current).
\alphasic{} was also measured at \SI{370}{\K}.
Data was captured using the \ibex{} software as configured for this instrument.
Measurements of an empty aluminium sample container were also taken at nominal temperatures \SI{18}{\K}, \SI{100}{\K}, \SI{200}{\K}, \SI{373}{\K}.

Direct-geometry \gls{ins} measurements were performed using the MARI neutron spectrometer at ISIS~\cite{leUpgradeMARISpectrometer2023}.
Each sample (\SI{25}{\gram} of \alphasic{} and \SI{20}{\gram} of \betagox{}) was loaded into an aluminium foil pouch of approximately $150 \times 40 \times 1$ \si{\milli\meter},
which was then rolled into an annulus and mounted in a cylindrical aluminium sample container in a helium \gls{ccr}.
A gadolinium Fermi chopper was used to select the incident neutron energy ($E_i$) giving resolution widths of around \SIrange{2}{4}{\percent} of $E_i$. Two settings were used for each sample: a low $E_i$, high resolution setting (\SI{60}{\milli\eV}, \SI{300}{\hertz} for \alphasic{} and \SI{120}{\milli\eV}, \SI{350}{\hertz} for \betagox{}) and a high $E_i$, high dynamic range setting (\SI{140}{\milli\eV}, \SI{450}{\hertz} for \alphasic{} and \SI{50}{\milli\eV}, \SI{200}{\hertz} for \betagox{}).
Measurements were taken with the sample temperature at \SI{5}{\K}, \SI{200}{\K} and \SI{450}{\K}, counting for $\sim$\SI{5}{\hour} (\SI{800}{\micro\ampere\hour} integrated proton current) at \SI{5}{\K} and proportionally less at higher temperatures. An empty aluminium foil in the sample container was also measured for approximately half that amount of time for background subtraction purposes.

\subsection{Phonon calculations} \label{sec:phonon-calcs}
First-principles phonon spectra were computed using density-functional perturbation theory and finite-displacement calculations implemented with \castep{}.

\subsubsection{\alphasic{}}
Structure data, determined by x-ray diffraction, was obtained from the ICSD for \sicTwoH{} (collection code 24261, $a=\SI{3.0763}{\angstrom}$, $c=\SI{5.0480}{\angstrom}$), \sicFourH{} (collection code 24170, $a=\SI{3.073}{\angstrom}$, $c=\SI{10.053}{\angstrom}$)
and \sicSixH{} (collection code 156190, $a=\SI{3.0810}{\angstrom}$, $c=\SI{15.1248}{\angstrom}$)%
~\cite{adamskySynthesisCrystallographyWurtzite1959,zagoracRecentDevelopmentsInorganic2019, thibaultMorphologicalStructuralCrystallography1944}.
Geometry optimisation and phonon calculations were performed across these three polytypes with (Perdew--Zunger) LDA, PBE, PBEsol and RSCAN \gls{xc} functionals~\cite{perdewSelfinteractionCorrectionDensityfunctional1981,perdewGeneralizedGradientApproximation1996,perdewRestoringDensityGradientExpansion2008,bartokRegularizedSCANFunctional2019}.
Additional calculations were performed in the 6H polytype using PBEsol with a D3 dispersion correction, including Becke--Johnson damping and three-body terms~\cite{mcnellisAzobenzeneCoinageMetal2009,grimmeConsistentAccurateInitio2010,grimmeEffectDampingFunction2011a}.

Atomic positions were first optimised using \castep{}~23 with \glspl{ncp} (and corresponding ``FINE" grid/cutoff settings), \SI{0.04}{\per\angstrom} \kpoint{} spacing and \SI{e-5}{\eV\per\angstrom} self-consistent-field criterion, until forces on all atoms were reduced below \SI{e-3}{\eV\per\angstrom}~\cite{clarkFirstPrinciplesMethods2005,refsonVariationalDensityfunctionalPerturbation2006}.
During structure relaxation, the lattice vectors were fixed according to the literature values while atomic positions were optimised within symmetry constraints.

A novel scheme was employed to ensure reciprocal-space convergence of the $\alpha$-\ce{SiC} phonons:
initial \gls{dfpt} calculations were performed with PBEsol in the \sicTwoH{} structure for an increasingly dense succession of \qpoint{} meshes,
and also at 16 quasi-random \qpoints{} obtained with the scrambled Sobol generator of \scipy{}~\cite{virtanenSciPyFundamentalAlgorithms2020,joeConstructingSobolSequences2008,owenScramblingSobolNiederreiter1998}.
The frequencies at quasi-random \qpoints{} were compared with values obtained by Fourier-interpolation of the regular meshes using \euphonic{},
and the residuals subjected to a basic statistical analysis.
A full comparison is shown in \suppl{} \cite{supp:material};
on a $5 \times 5 \times 3$ mesh the standard deviation of error falls below \SI{1}{\per\cm}, and this sampling was selected for all \gls{ins} simulations of \sicTwoH{}.
It was assumed that this \qpoint{} spacing would translate to equivalent supercells in finite-displacement real-space phonon calculations, and when using other XC functionals.
For the longer \sicFourH{} and \sicSixH{} unit cells, equivalent-or-better $5\times5\times2$ and $5\times5\times1$ \qpoint{} meshes were used respectively.

As \gls{dfpt} was not available for meta-GGA or D3-corrected calculations in \castep{}, RSCAN and PBEsol-D3 calculations were performed using the efficient finite-displacement scheme of Monserrat and Lloyd-Williams~\cite{lloyd-williamsLatticeDynamicsElectronphonon2015}.
To compensate for the increased cost of RSCAN calculations the ``C19'' ultrasoft pseudopotential set was used: for \sicTwoH{} with a ``FINE'' plane-wave cutoff of \SI{\sim 353}{\eV} and corresponding default grids,
while for \sicFourH{} and \sicSixH{} this was tightened to a \SI{550}{\eV} plane-wave cutoff and \SI{42}{\per\angstrom} fine grid to eliminate some initially-observed soft modes.
To account for long-range dipole-dipole interactions, RSCAN and PBEsol-D3 force constants were combined with Born effective charges and dielectric tensor from PBEsol \gls{dfpt} calculations using \castep{}~25.

\subsubsection{\betagox{}}
A similar approach was employed for convergence of \betagox{} \gls{dfpt} calculations with PBEsol.
The experimental structure was retrieved from the ICSD (collection code 83645)~\cite{zagoracRecentDevelopmentsInorganic2019,ahmanReinvestigationVGalliumOxide1996}.
Initial ``singlepoint'' \gls{dft} calculations with PBEsol and \gls{ncp} were used to select a plane-wave cutoff of \SI{1200}{\eV}, appearing to give force convergence of \SI{< 10}{\meV\per\angstrom}.

A Minkowski-reduced primitive cell was derived from this using the Atomic Simulation Environment: the lattice parameters are $\alpha=\SI{103.41}{\degree}, \beta=\SI{103.96}{\degree},  \gamma=\SI{90.00}{\degree}$, $a=\SI{3.0371}{\angstrom}, b=\SI{5.7981}{\angstrom}, c=\SI{6.2930}{\angstrom}$~\cite{nguyenLowdimensionalLatticeBasis2009,hjorthlarsenAtomicSimulationEnvironment2017}.
A \kpoint{} spacing of \SI{0.05}{\per\angstrom} was selected, giving total energy convergence of the order \SI{1}{\meV}.
Geometry optimisation was performed to reduce forces below \SI{1e-4}{\eV\per\angstrom}, with fixed lattice parameters.
A $2\times3\times2$ supercell of this primitive cell was found to adequately match the frequencies of quasirandom \qpoint{}{} samples when Fourier-interpolated,
giving an error distribution with standard deviation below \SI{1.5}{\per\cm}.
Optimisation and phonon calculations were repeated with the LDA and RSCAN \gls{xc} functionals.
Again, RSCAN calculations used a finite-displacement scheme to obtain the force constants using C19 pseudopotentials and were combined with PBEsol/\gls{dfpt} data for long-range interactions;
this time the basis-set cutoff was set to \SI{700}{\eV} and fine integration grid cutoff to \SI{42}{\per \angstrom}.

\subsubsection{Al}
Phonons for the aluminium sample container were calculated within the \gls{lda} with \gls{dfpt}/\gls{ncp} as this has previously shown good agreement with experiment for \gls{ins} simulations using \euphonic{}~\cite{fairEuphonicInelasticNeutron2022a}.
The primitive cell of FCC Al was used with lattice parameter $a=\SI{4.0323}{\angstrom}$;
this was assumed to be representative at low temperature, as a rough median value between reported temperature--$a$ relationships~\cite{straumanisLatticeParametersThermal1971b}.
\glspl{ncp} were used with a \kpoint{} spacing of \SI{0.022}{\per\angstrom} (converging total energy to $\sim\SI{1}{\meV}$ per atom) and \SI{450}{\eV} plane-wave cutoff (converging forces of $\sim\SI{0.7}{\eV\per\angstrom}$ in a distorted structure to $\sim\SI{1}{\meV\per\angstrom}$).
\qpoint{} meshes were tested from $2 \times2 \times2$ up to $7\times 7 \times 7$; the largest mesh gives errors with a standard deviation below \SI{1}{\per\cm} and is used in this work.

\subsection{Simulated INS spectra}
\subsubsection{Incoherent approximation}
The calculated force constants were analyzed with the \abins{} routine included in \mantid{} version 6.12. This makes an incoherent \gls{dos}-like approximation as described in \cref{sec:sim-methods-intro}:
the whole Brillouin zone is sampled using \euphonic{} to obtain frequencies and eigenvectors,
instrument constraints are considered and the intensity is calculated along the
appropriate $(\omega, \mathbf{q})$ line (for TOSCA, with the ``Abins" algorithm)
or in an accessible $(\omega, \mathbf{q})$ region (for MARI, with the ``Abins2D" algorithm).

\subsubsection{Numerical sampling of coherent spectrum}
A Python/Snakemake workflow has been written, making use of recent improvements to the \euphonic{} library~\cite{molderSustainableDataAnalysis2025,fairEuphonicInelasticNeutron2022a}.
The coherent scattering spectrum is calculated in ``shells'' of quasi-random \qpoints{} to obtain a numerical powder-averaged spectrum in $S(Q, \omega)$.
The $S(Q, \omega)$ lines accessible to TOSCA are drawn from this grid and averaged to make a 1-D spectrum for each of the two detector banks.
These are then broadened with the same energy-dependent Gaussian resolution model as in \abins{}.
MARI measurements are also simulated at the relevant temperatures and energy ranges by masking out the inaccessible $S(Q, \omega)$ regions and broadening with an energy-dependent Gaussian FWHM from \codename{pychop}~\cite{arnold_mantiddata_2014}.

The spectrum was calculated in quasi-random spherical shells of \num{8 000} \qpoints{} at \SI{0.025}{\per\angstrom} intervals, with \SI{0.1}{\per\angstrom}-FWHM Gaussian broadening applied along the $Q$ axis.
The coherent scattering calculations with \euphonic{} included thermal occupation and Debye--Waller factors, using displacement vectors computed on a unit-cell-dependent \qpoint{} mesh with spacing below \SI{0.1}{\per\angstrom}.

\section{Results}
\subsection{TOSCA measurements}\label{sec:results-tosca}
Data from TOSCA was reduced to $S(\omega)$ plots using the \mantid{} software framework.
Results are given in \si{\barn \cm} to reflect the nature of $S$ as a spectrum of scattering area;
however an unknown scale factor is present due to sample size and instrument efficiency.
(In this work the simulation codes have their own reference units and the output is roughly scaled where necessary to aid comparisons.
The scaling is made consistently for each sample/instrument combination.)
Reference measurements of an empty aluminium sample-holder were subtracted from the TOSCA spectra.
These subtractions are shown in the SI: the general effect is to remove some broad peaks from \SIrange{100}{200}{\per\cm} and a flat temperature-insensitive background. In the \betagox{} case the background intensity is quite significant compared to the signal, making it more difficult to distinguish between noise and small peaks.

\begin{figure}
    \centering
    \includegraphics{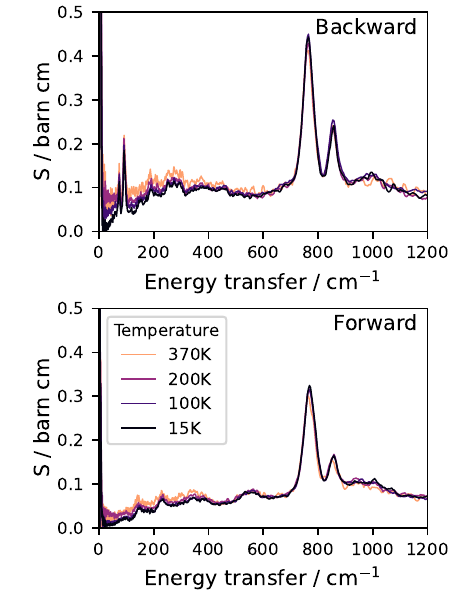}
    \caption{Temperature trend in TOSCA SiC measurements. Data has been background-subtracted and Wiener filtering applied to aid comparison of main features.}
    \label{fig:sic-temperature}
\end{figure}

\begin{figure}
    \centering
    \includegraphics{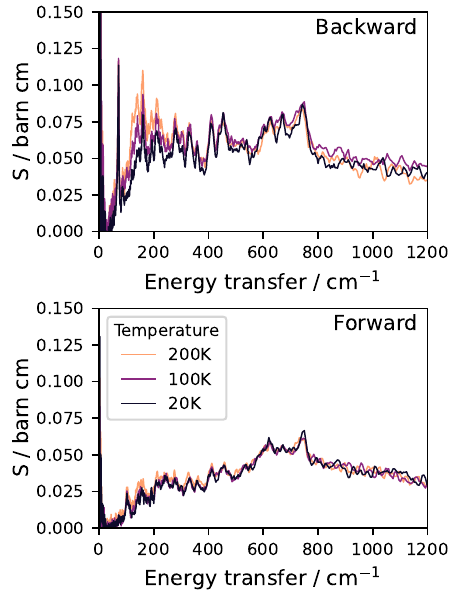}
    \caption{Temperature trend in TOSCA \ce{Ga2O3} measurements. Data has been background-subtracted and Wiener filtering applied to aid comparison of main features.}
    \label{fig:ga2o3-temperature}
\end{figure}

The effect of temperature on the resulting spectra is illustrated in \cref{fig:sic-temperature,fig:ga2o3-temperature}; to aid comparison in these figures, they are de-noised with a Wiener filter as implemented in \scipy{}~\cite{virtanenSciPyFundamentalAlgorithms2020}.
We see that for both ceramics the peak positions are insensitive to temperature in this range,
and the overall intensity exhibits only a slight ``clockwise tilt'' at high temperature,
as thermal occupation increases at low frequencies and a stronger Debye-Waller factor dampens the high-frequency spectrum.

In each spectrum there are some sharp peaks below \SI{100}{\per\cm} which are not considered to be part of the powder phonon spectrum.
These only appear in the backward-scattering detector bank over a subset of the detectors (i.e. associated with some range of angles $\varphi$ orthogonal to the scattering angle $\theta$.)
They are assumed to be an artifact of the experimental setup.

\subsection{Coherent-scattering simulations} \label{sec:slice-sims}
\begin{figure*}

\subfloat[Aluminium spectrum simulated with \castep{}/LDA/\gls{dfpt}]{
    \includegraphics[width=0.98\textwidth]{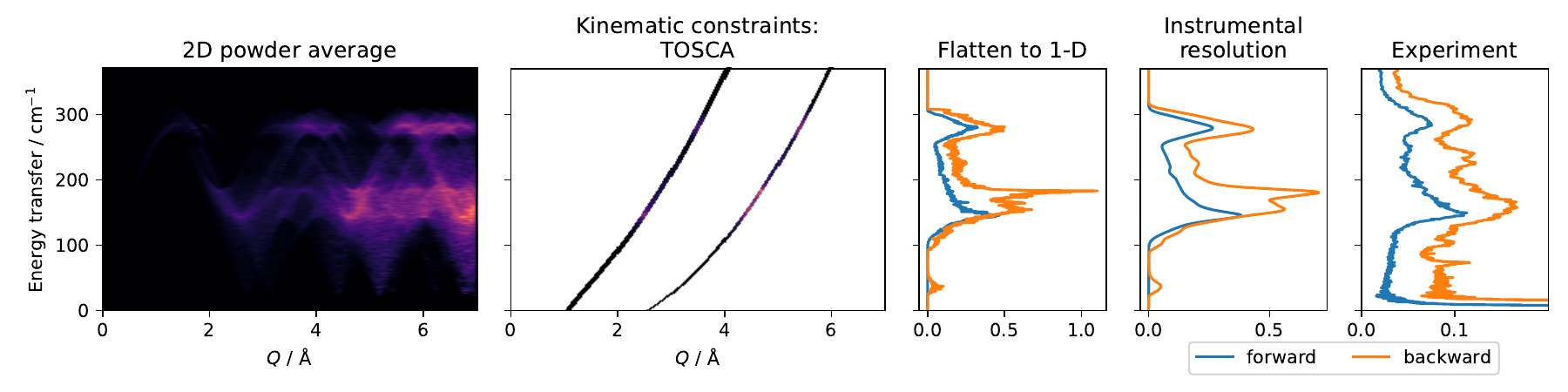}
    \label{fig:sliced-al}
}

\subfloat[\alphasic{} measurement simulated with \castep{}/PBEsol/DFPT in 6H polytype]{
    \includegraphics[width=0.98\textwidth]{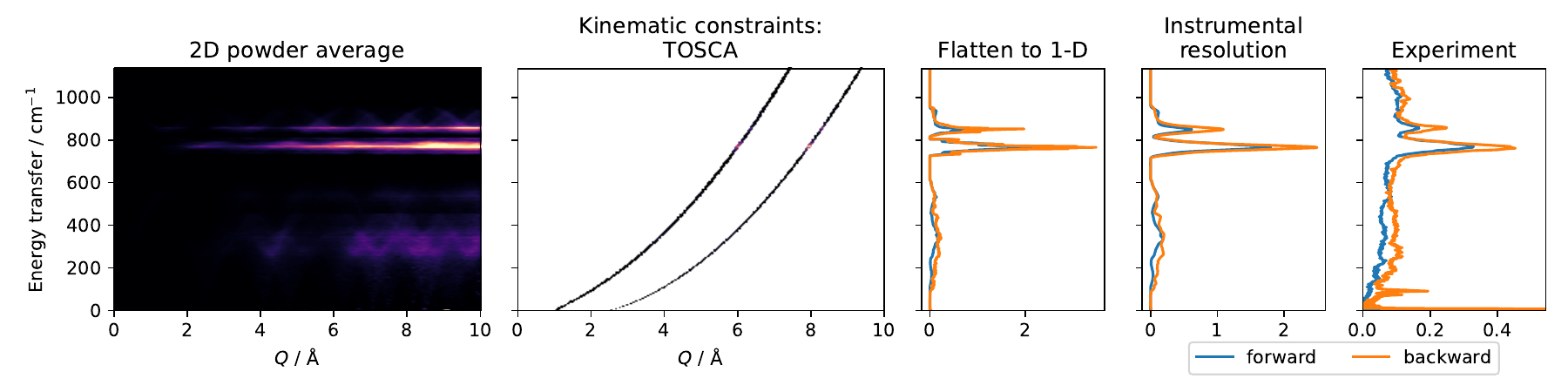}
    \label{fig:sliced-sic}
}

\subfloat[\betagox{} measurement simulated with \castep{}/RSCAN]{
    \includegraphics[width=0.98\textwidth]{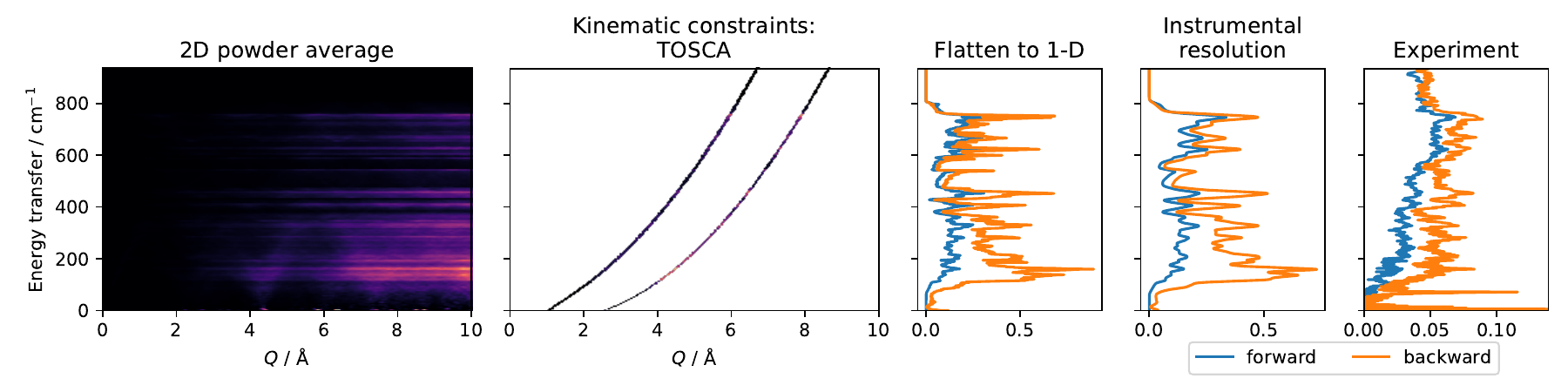}
    \label{fig:sliced-ga2o3}
}

\caption{Simulated TOSCA measurements: coherent scattering in powder average is sliced along the $(\omega, Q)$ constraints of each detector bank, and broadened with instrumental resolution function $\sigma(\omega)$.}
\label{fig:sliced}
\end{figure*}

The spectra at forward/backward detectors for low-temperature measurements are compared (without smoothing) to the \euphonic{} simulations in \cref{fig:sliced}.
In the aluminium case we find that the simulation with \gls{lda} has generally captured the peak frequencies and the difference between forward/backward detector banks. The most notable differences are:
\begin{itemize}
    \item The experimental peak at \SI{280}{\per\cm} is broader and weaker than the simulation.
    \item Peaks from \SIrange{200}{250}{\per\cm} are weaker in the simulation.
    \item In the back-scattering spectrum the simulated intensity around \SI{180}{\per\cm} is higher than expected, giving the appearance of split peaks rather than one broad peak.
\end{itemize}

For \alphasic{} the best agreement in optic mode frequency and shape was found in the \sicSixH{} simulations using PBEsol;
this is consistent with the \gls{xrd} analysis.
The calculated phonon band structures and \gls{ins} simulations with all the considered polytypes and \gls{xc} functionals are compared in the \suppl{} \cite{supp:material}.
Summarising the observations:
the 2H polytype calculations reveal more peak-splitting in optic modes than the experimental spectrum,
while the 4H structure gives fair agreement with experiment spectra especially when using \gls{lda};
RSCAN was found to give generally higher frequencies than the \gls{gga} functionals, with the \gls{lda} spectrum red-shifted relative to the experimental measurements;
the PBEsol-D3 results were indistinguishable from the uncorrected PBEsol for \sicSixH{}.
The calculations in \sicSixH{} polytype with PBEsol are used for all the relevant figures in this manuscript.
In \cref{fig:sliced-sic} we see good agreement in the optic modes and a matching relationship between forward/backward scattering spectra.
Again the experimental peaks are a bit broader than simulated.

In the \betagox{} phonon calculations, a similar trend of increasing frequencies was observed from LDA to GGA to RSCAN;
in this case the result is that RSCAN gives the best agreement with the experimental spectra.
The most prominent discrepancy in \cref{fig:sliced-ga2o3} is that the intensity around \SI{160}{\per\cm} is overestimated in simulation;
this might be related to lifetime broadening from phonon-phonon interactions, which is neglected in the harmonic approximation.
(As discussed above, the sharp peak at \SI{72}{\per\cm} in experiment is assumed to be an artifact of the TOSCA measurement.)

\subsection{MARI measurements}
\begin{figure*}
    \subfloat[MARI measurement, $E_i = \SI{60}{\meV}$]{\includegraphics[width=0.48\textwidth]{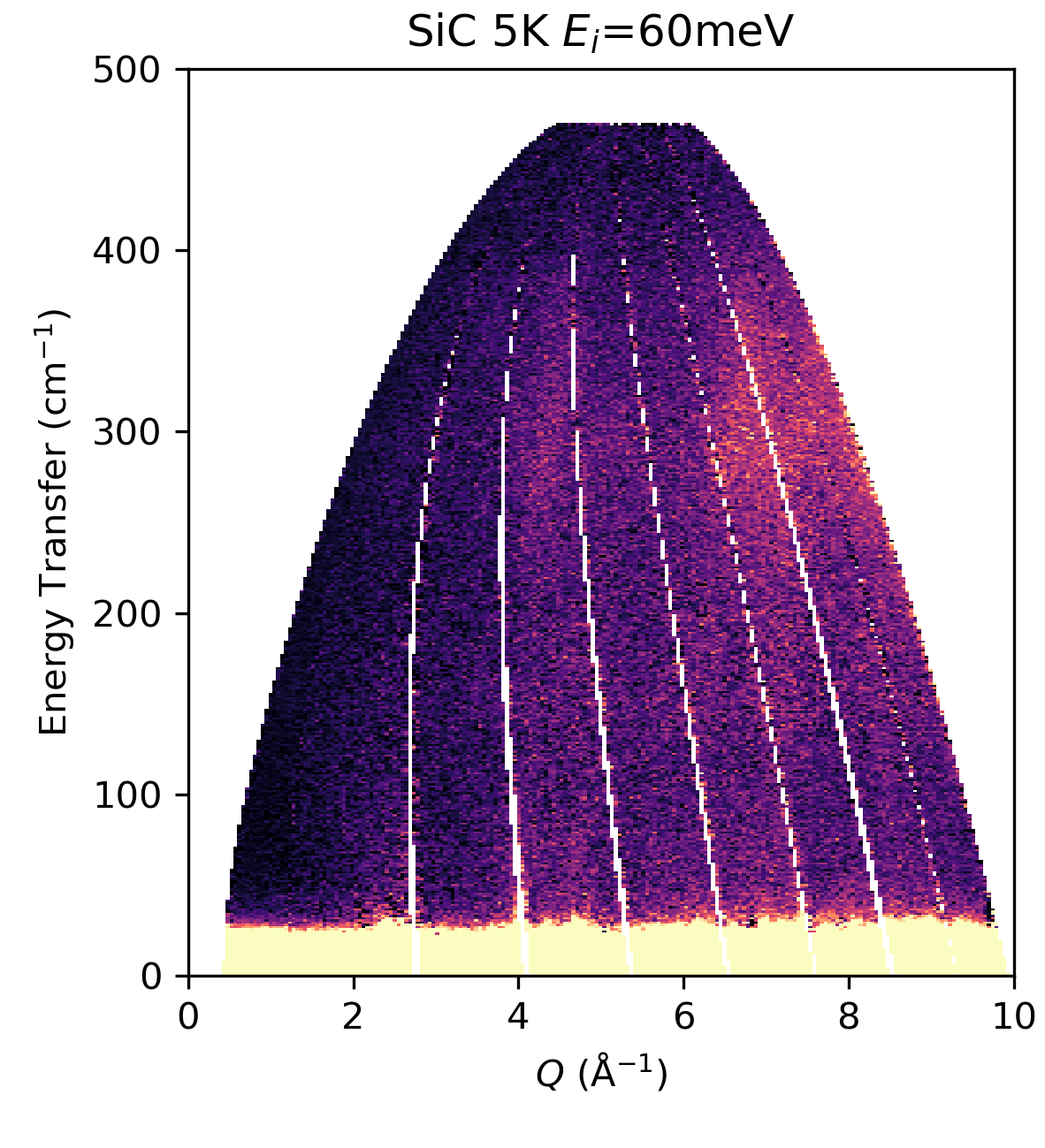}
    \label{subfig:mari-sic-5k-expt}
    }
    \subfloat[Simulation]{
    \includegraphics[width=0.48\textwidth]{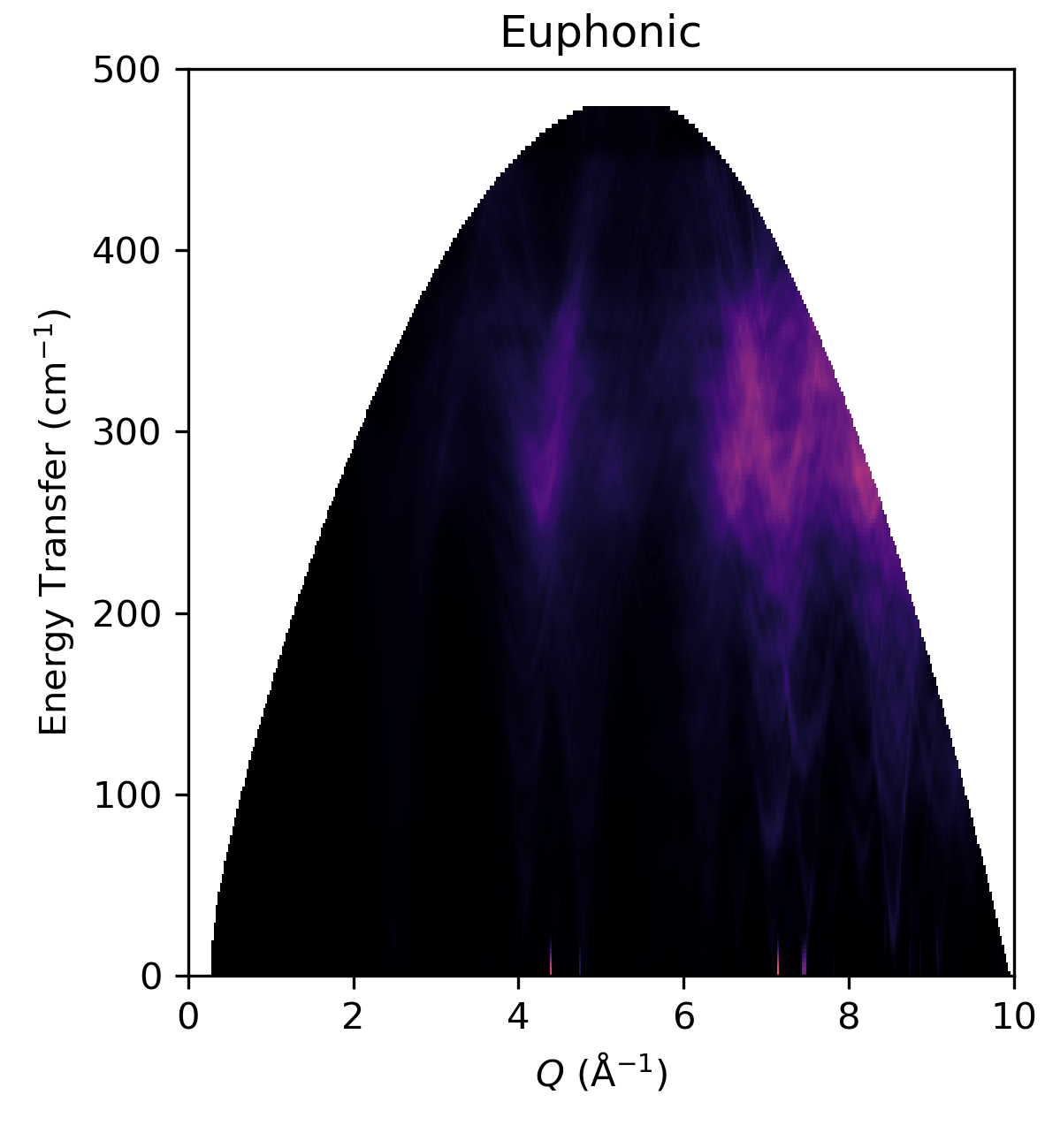}
    \label{subfig:mari-sic-5k-sim}
}
    \\
    \subfloat[MARI measurement, $E_i = \SI{140}{meV}$]{
    \includegraphics[width=0.48\textwidth]{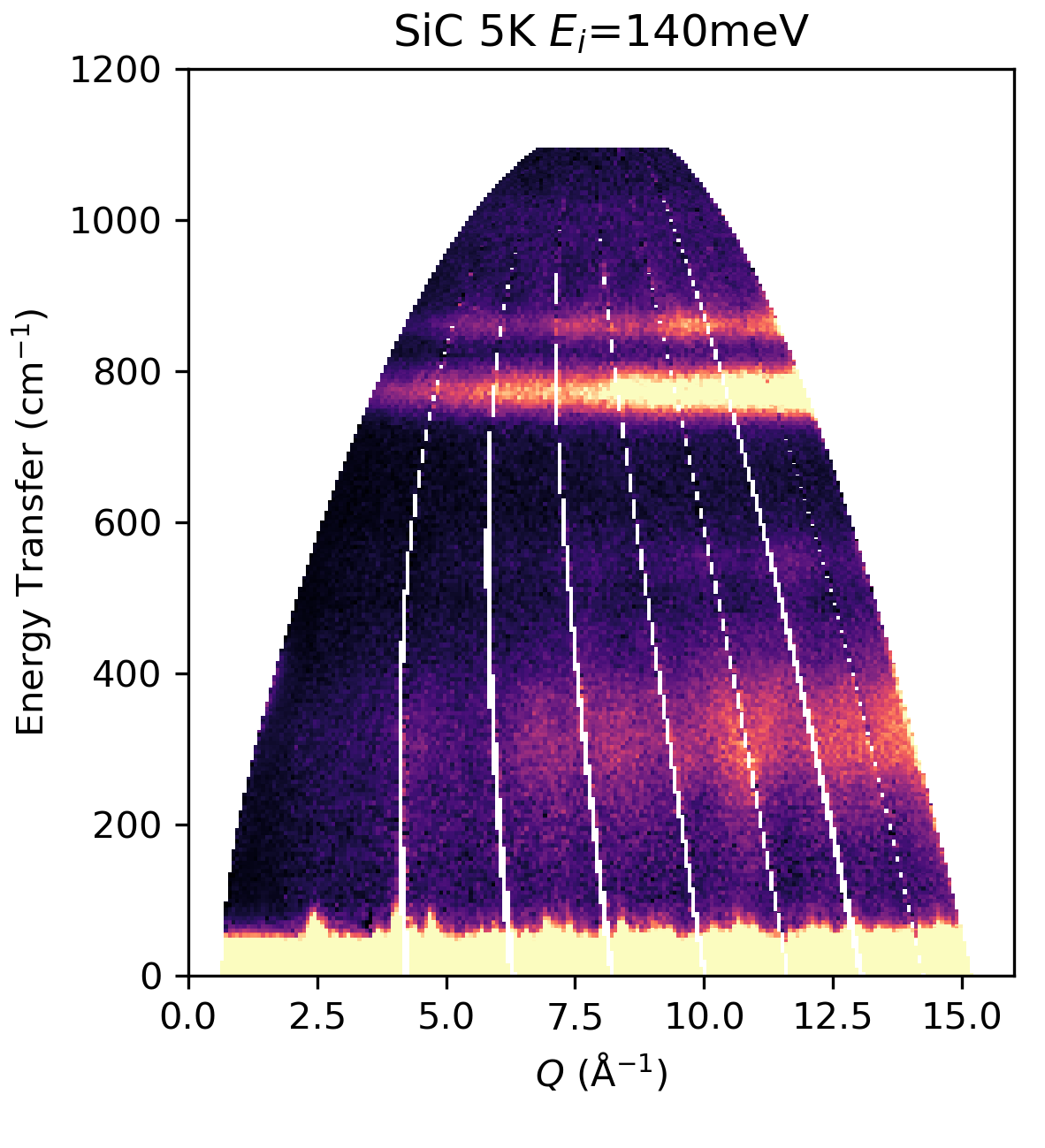}
    \label{subfig:mari-sic-5k-140-expt}    
    }
    \subfloat[Simulation]{
    \includegraphics[width=0.48\textwidth]{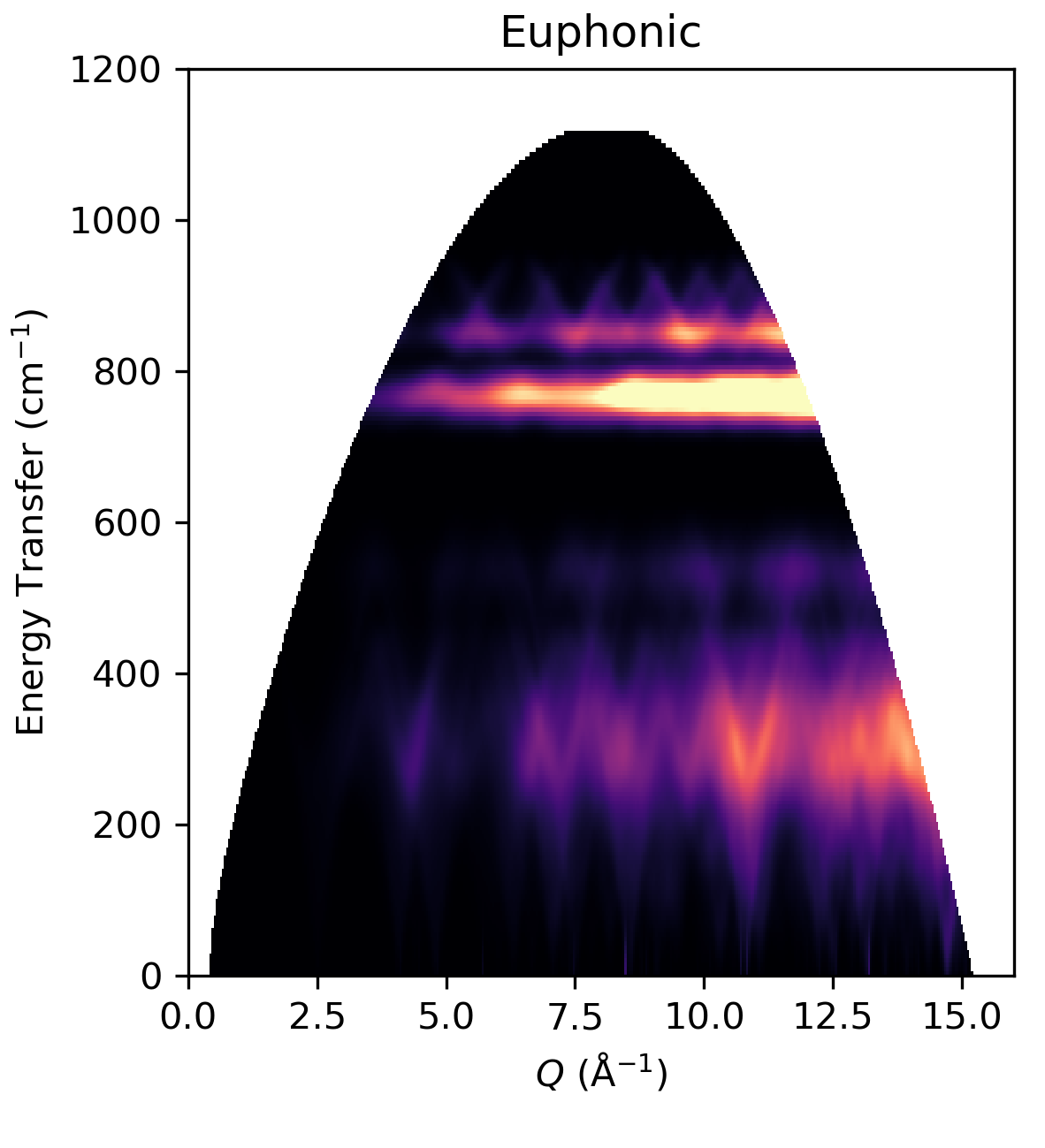}
    \label{subfig:mari-sic-5k-140-sim}    
    }   
    \caption{
    \ce{SiC} $S(Q, \omega)$ maps measured on MARI at nominal temperature 5K with \SI{60}{\meV} incident energy, \SI{200}{\Hz} chopper \protect\subref{subfig:mari-sic-5k-expt} and \SI{140}{\meV} at \SI{450}{\Hz} \protect\subref{subfig:mari-sic-5k-140-expt} and corresponding simulations (\protect\subref*{subfig:mari-sic-5k-sim}), (\protect\subref*{subfig:mari-sic-5k-140-sim}).
    }
    \label{fig:mari-sic-5K}
\end{figure*}

\begin{figure*}
\subfloat[MARI measurement, $E_i = \SI{50}{\meV}$]{
    \includegraphics[width=0.48\textwidth]{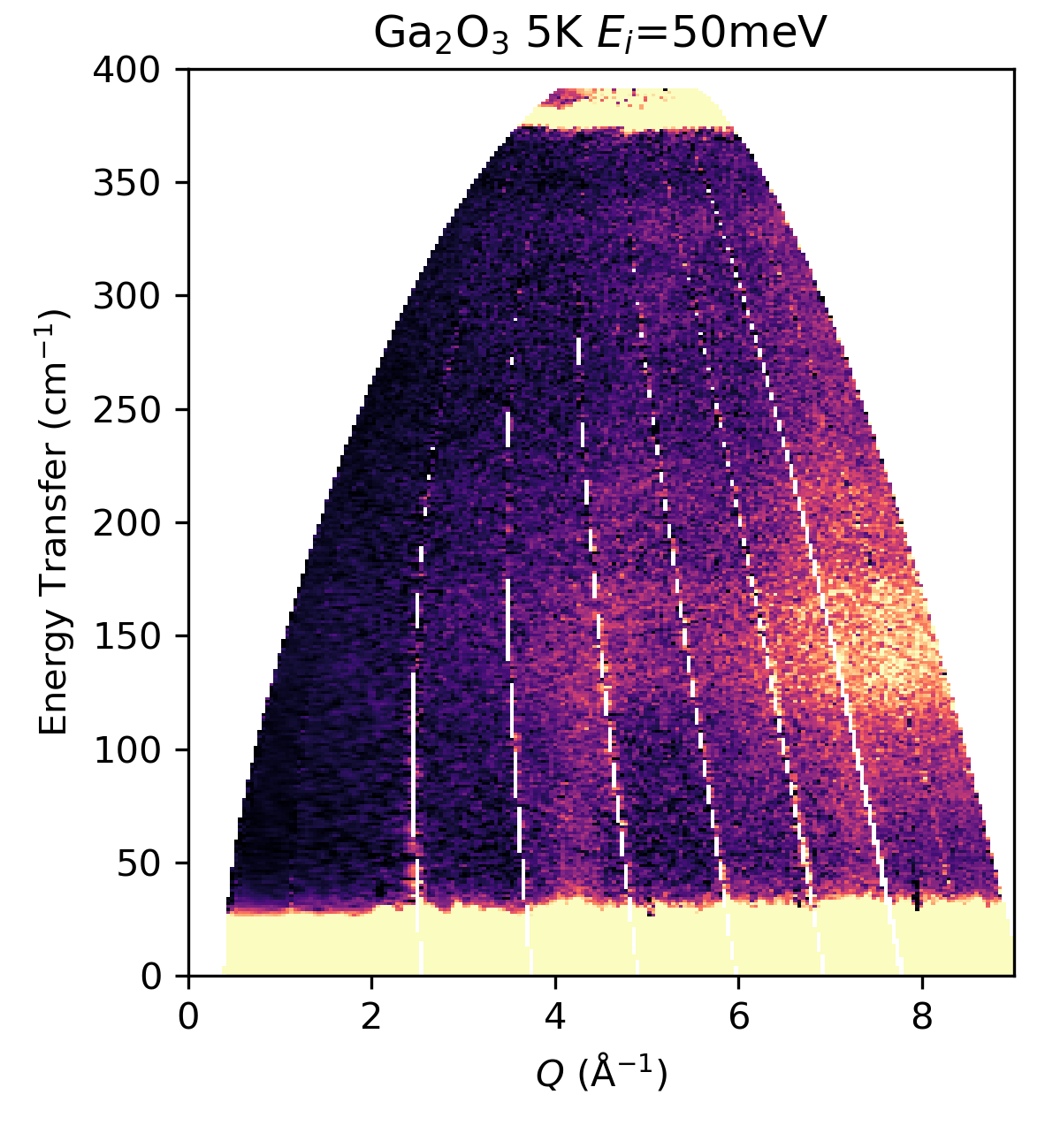}
    \label{subfig:mari-ga2o3-5k-expt}
}
\subfloat[Simulation]{
    \includegraphics[width=0.48\textwidth]{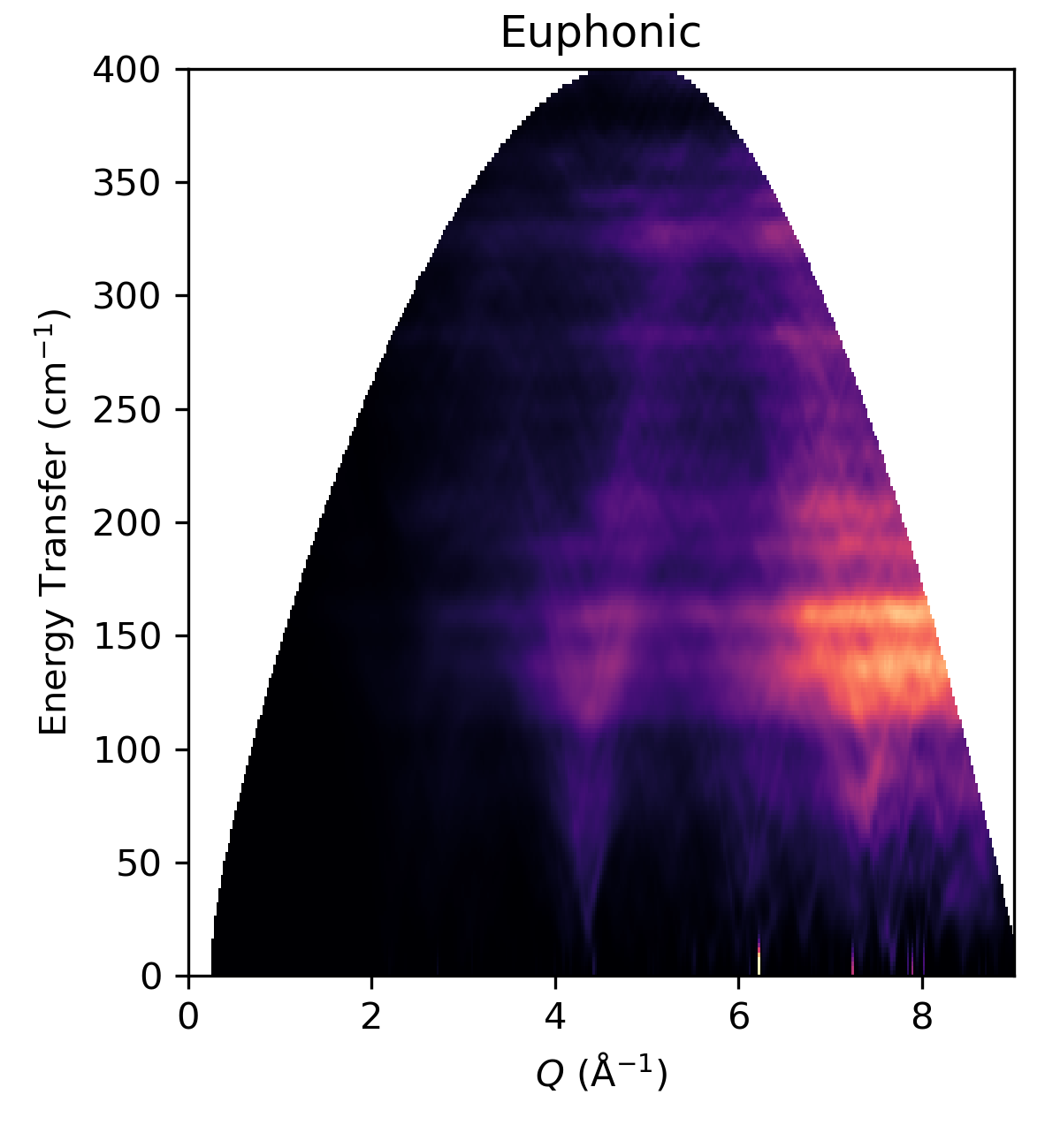}
    \label{subfig:mari-ga2o3-5k-sim}}
\\
\subfloat[MARI measurement, $E_i = \SI{120}{meV}$]{
    \includegraphics[width=0.48\textwidth]{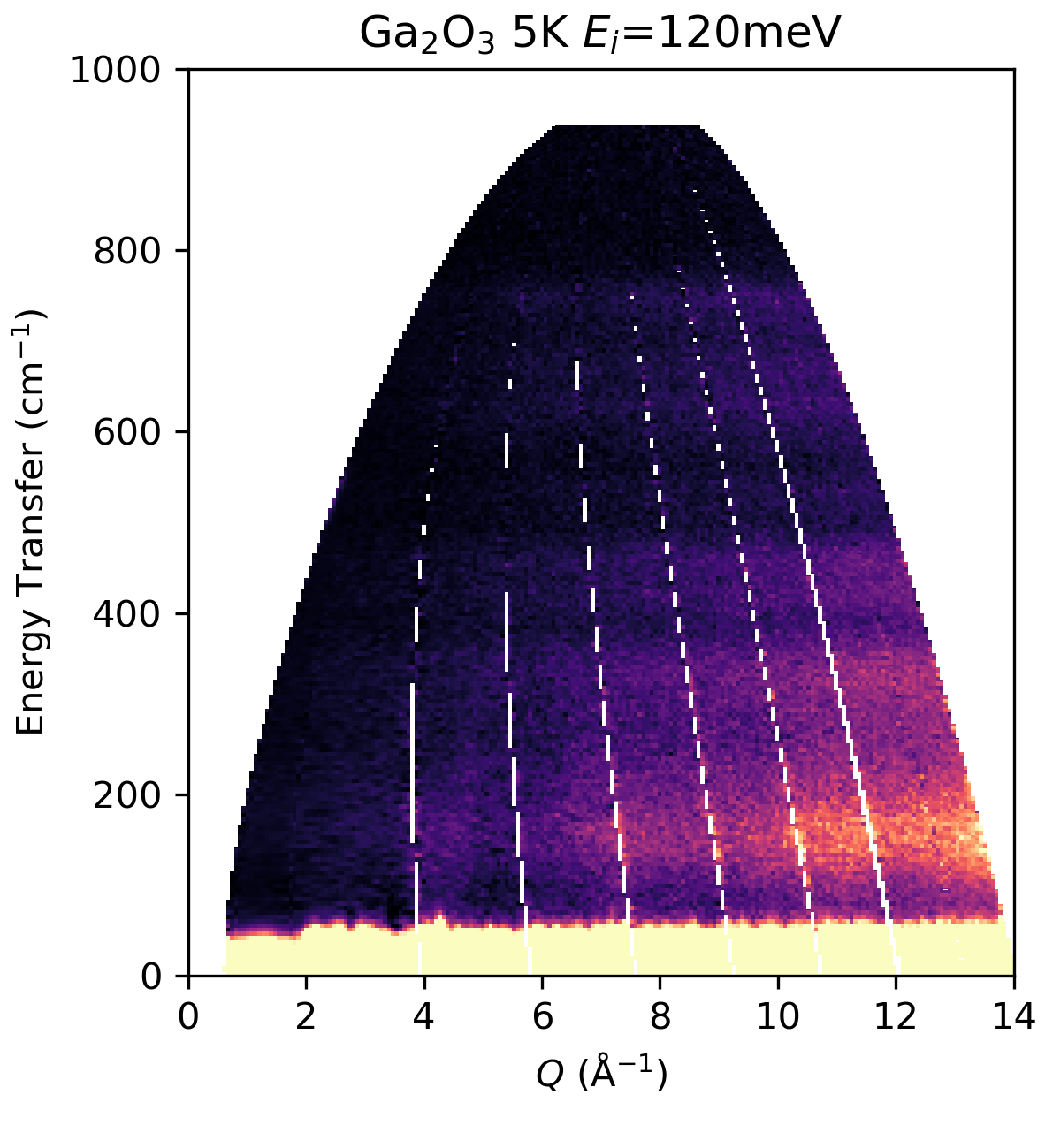}
    \label{subfig:mari-ga2o3-5k-120-expt}
}
\subfloat[Simulation]{
    \includegraphics[width=0.48\textwidth]{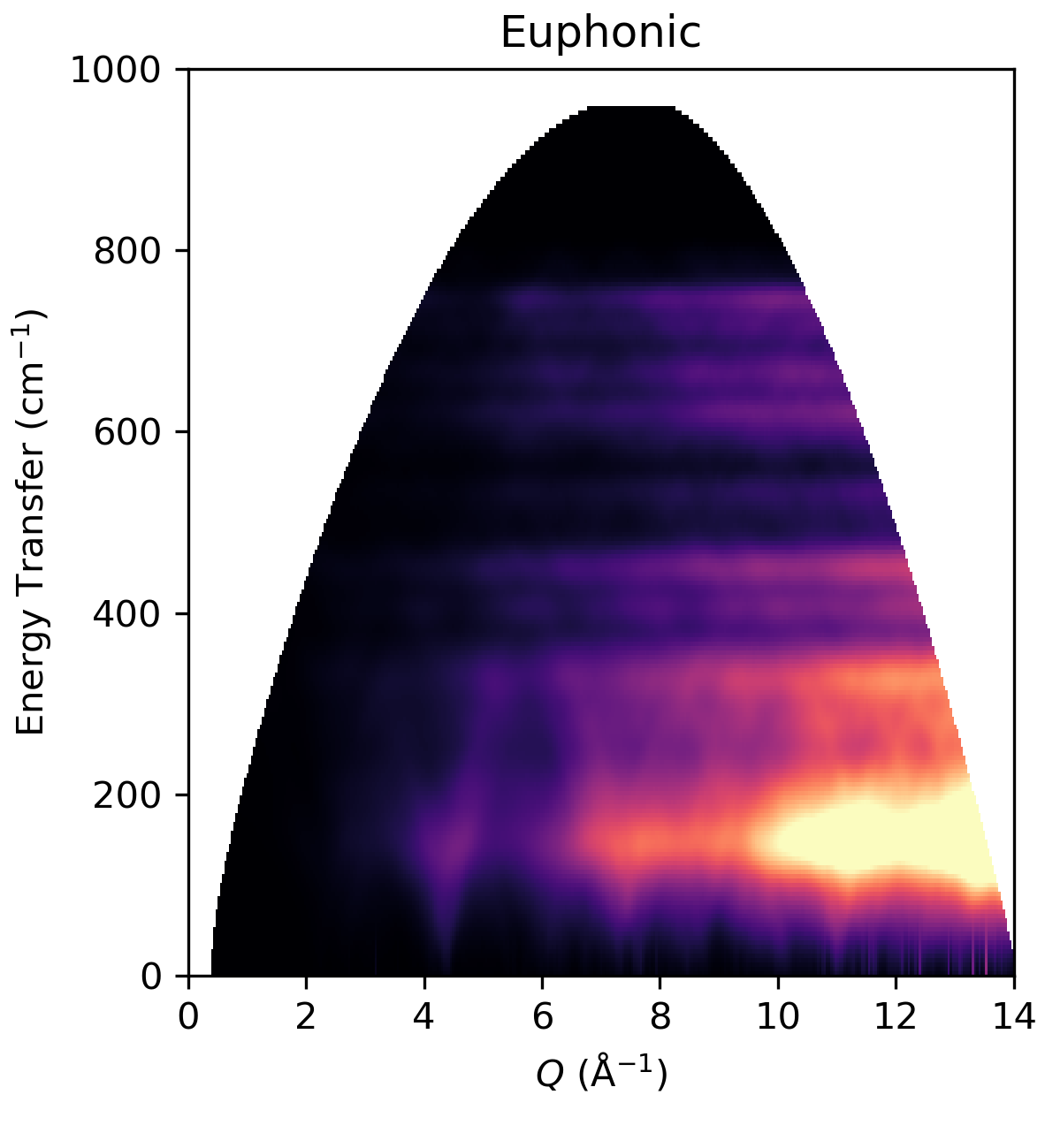}
    \label{subfig:mari-ga2o3-120-5k-sim}
}

    \caption{
    \ce{Ga2O3} $S(Q, \omega)$ maps measured on MARI at nominal temperature 5K with \SI{50}{\meV} incident energy,
    \SI{200}{\Hz} chopper \protect\subref{subfig:mari-ga2o3-5k-expt} and
    \SI{120}{\meV} at \SI{350}{\Hz} (\protect\subref{subfig:mari-ga2o3-5k-120-expt}), and corresponding simulations (\protect\subref*{subfig:mari-ga2o3-5k-sim}, \protect\subref*{subfig:mari-ga2o3-120-5k-sim}).
    \label{fig:mari-ga2o3-5K}
    }
\end{figure*}

The most feature-rich and low-noise MARI measurements are shown with corresponding powder-averaged coherent-scattering simulations in \cref{fig:mari-sic-5K,fig:mari-ga2o3-5K}. A full set of $S(Q, \omega)$ maps at different temperatures and incident energies is included in \suppl{} \cite{supp:material}.
We find that in general the agreement between theory and experiment is satisfactory,
validating the powder-averaged fundamental coherent phonon scattering simulation as
a model of \gls{ins} from these samples
and as the basis for TOSCA simulations in \cref{sec:slice-sims}.
The bright horizontal band at around \SI{380}{\per\cm} in (\cref{subfig:mari-ga2o3-5k-expt}) is assigned to neutrons from the $E_i=\SI{16}{\meV}$ ``rep" of this multiple-incident-energy measurement.
Other than this, the most striking difference between experiment and simulation is the bright elastic-scattering feature close to zero energy transfer; this is not included in the inelastic scattering simulation.
The 2-D masking approach also neglects small gaps in $S(Q,\omega)$ coverage between detector banks.
Further assignment of ``spurion'' features is made in the \suppl{} \cite{supp:material}.

\subsection{Incoherent approximation simulations}
\begin{figure}
    \subfloat[]{\includegraphics[width=0.5\textwidth]{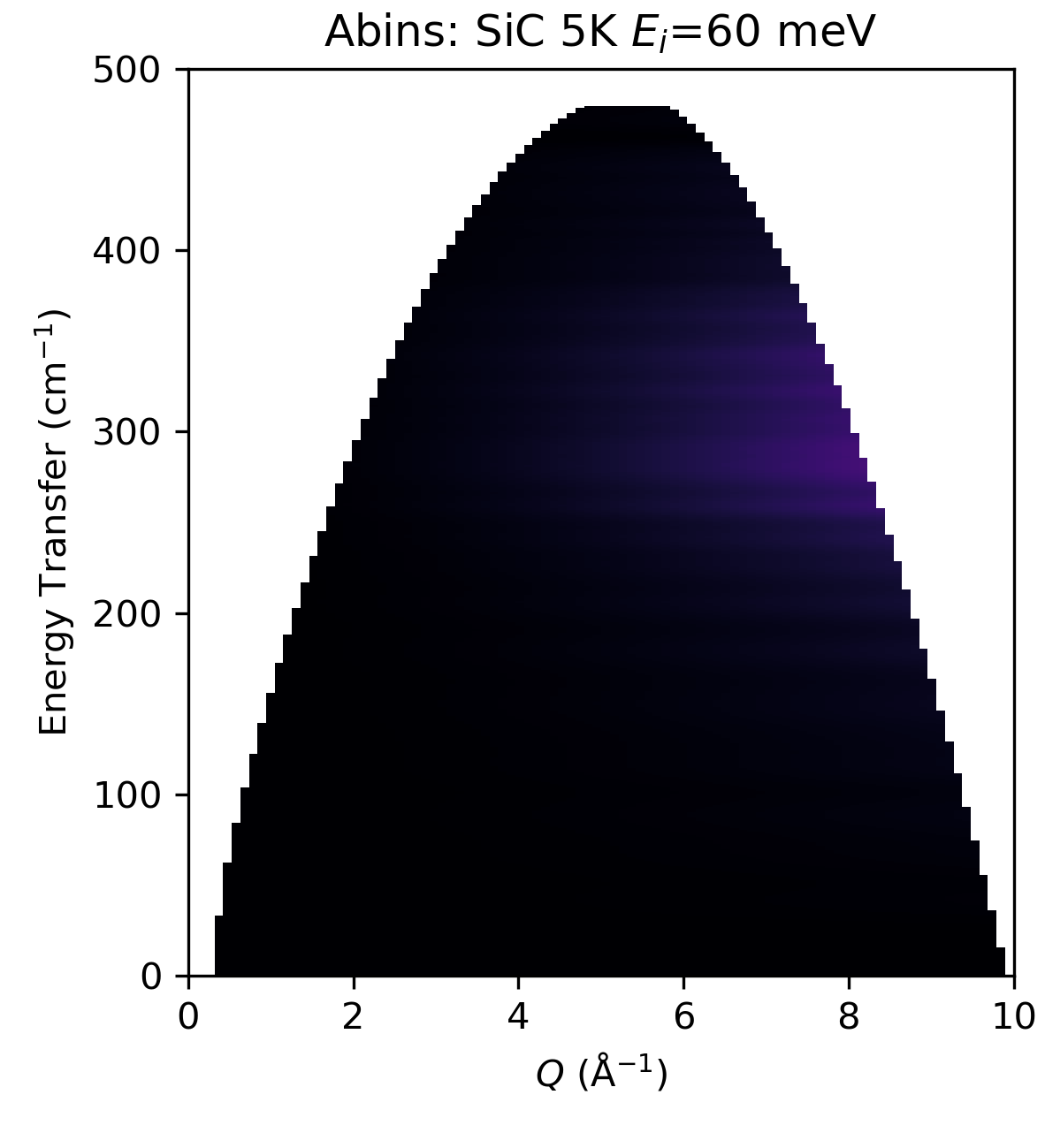}}
    
    \subfloat[]{\includegraphics[width=0.5\textwidth]{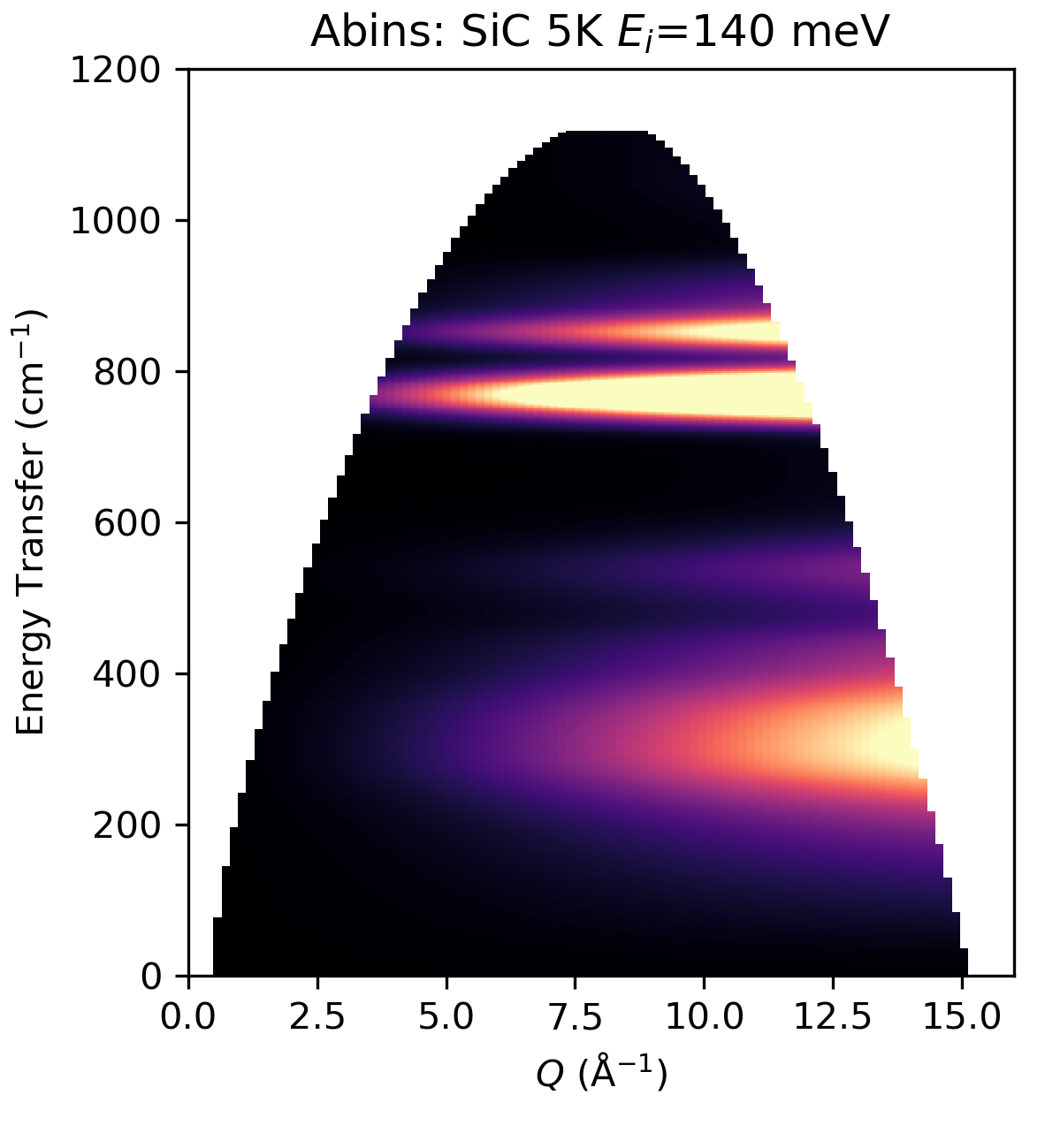}}
    \caption{\abins{} simulations of MARI \alphasic{} spectra in incoherent \gls{dos}-like approximation}
    \label{fig:sic-abins}
\end{figure}

\begin{figure}
    \subfloat[]{\includegraphics[width=0.5\textwidth]{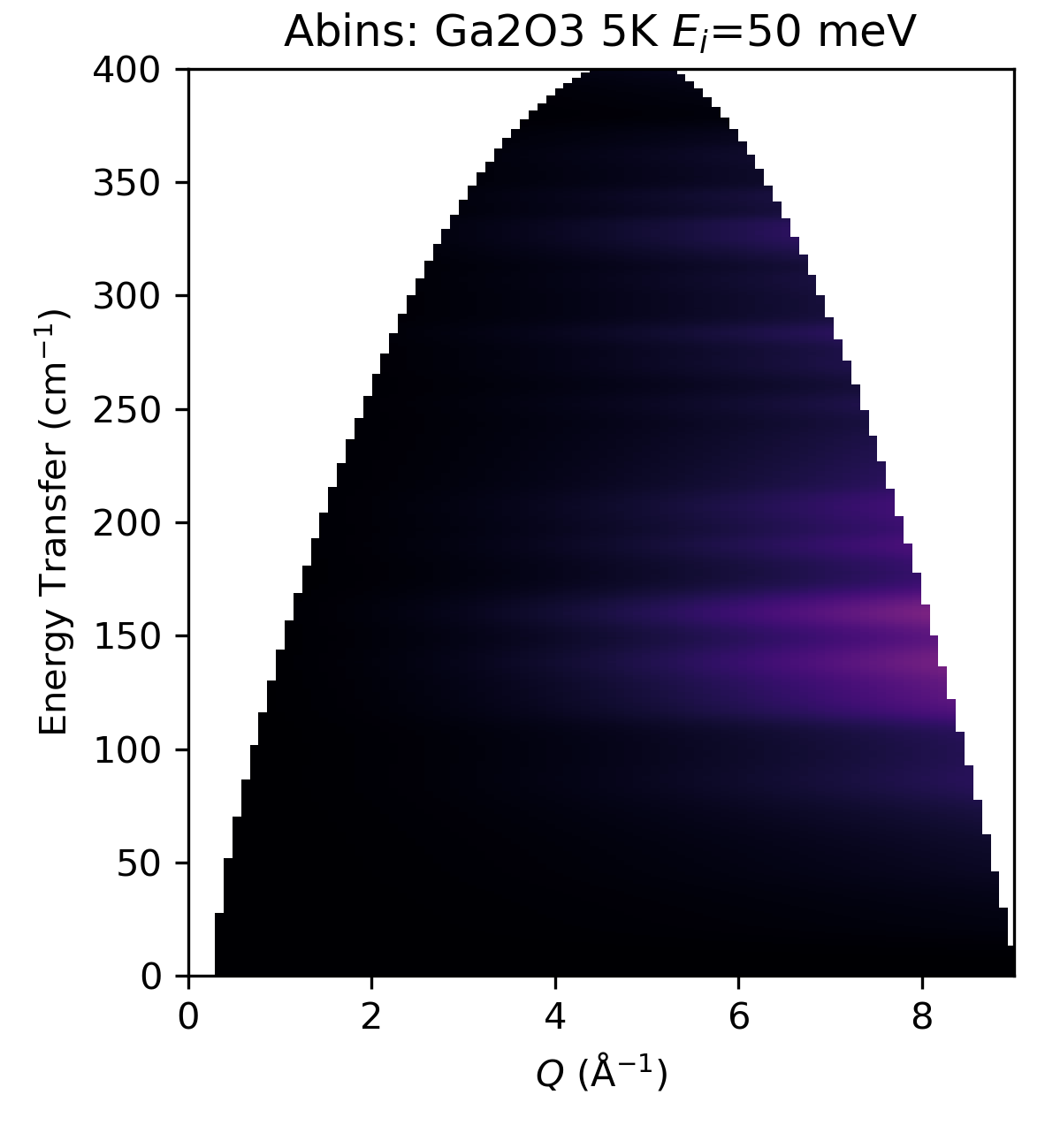}}

    \subfloat[]{
        \includegraphics[width=0.5\textwidth]{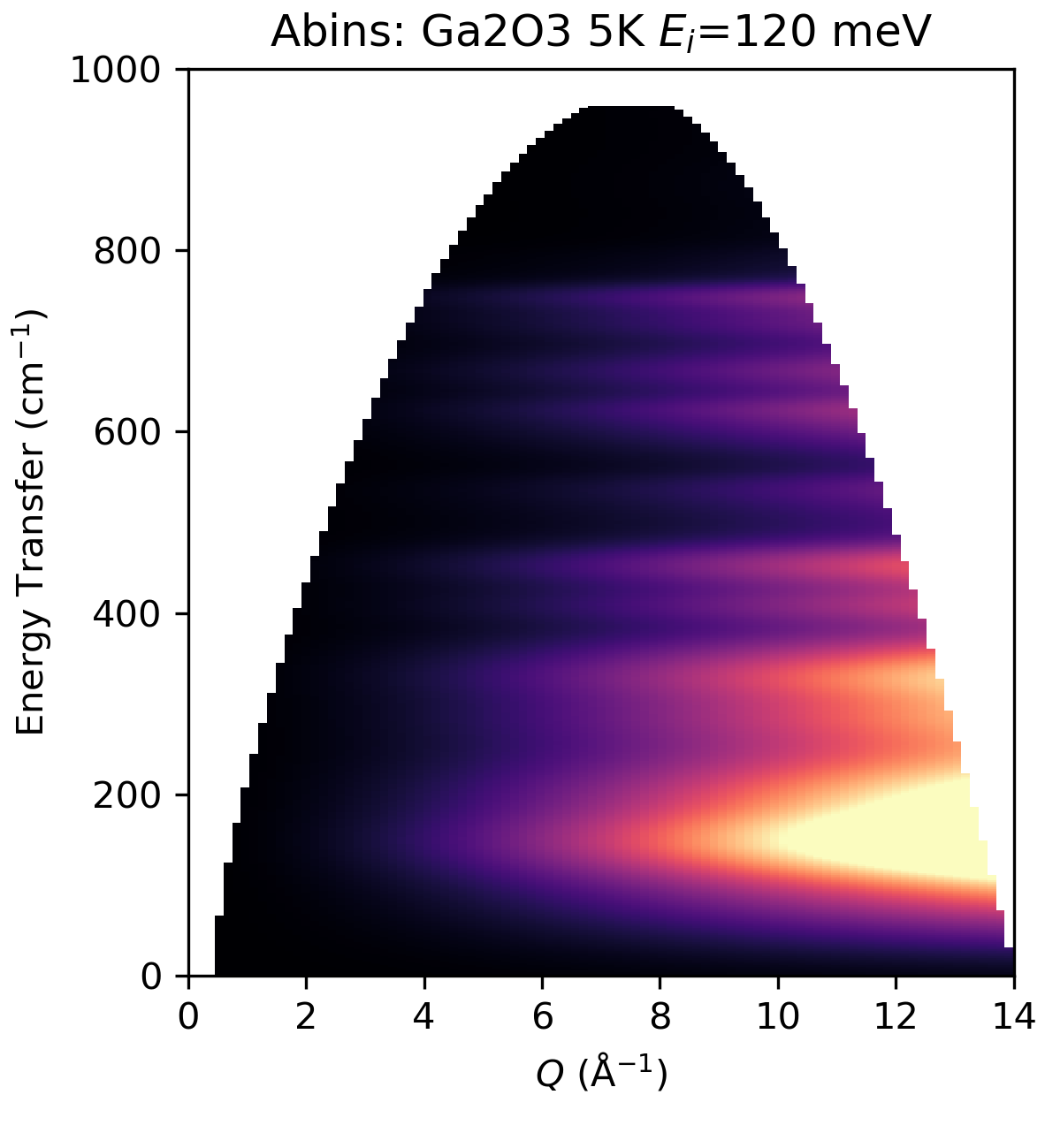}
}
    \caption{\abins{} simulations of MARI \betagox{} spectra in incoherent \gls{dos}-like approximation}
    \label{fig:ga2o3-abins}
\end{figure}

\begin{figure}
    \subfloat[Aluminium (LDA calculations)]{
        \includegraphics[width=0.5\textwidth]{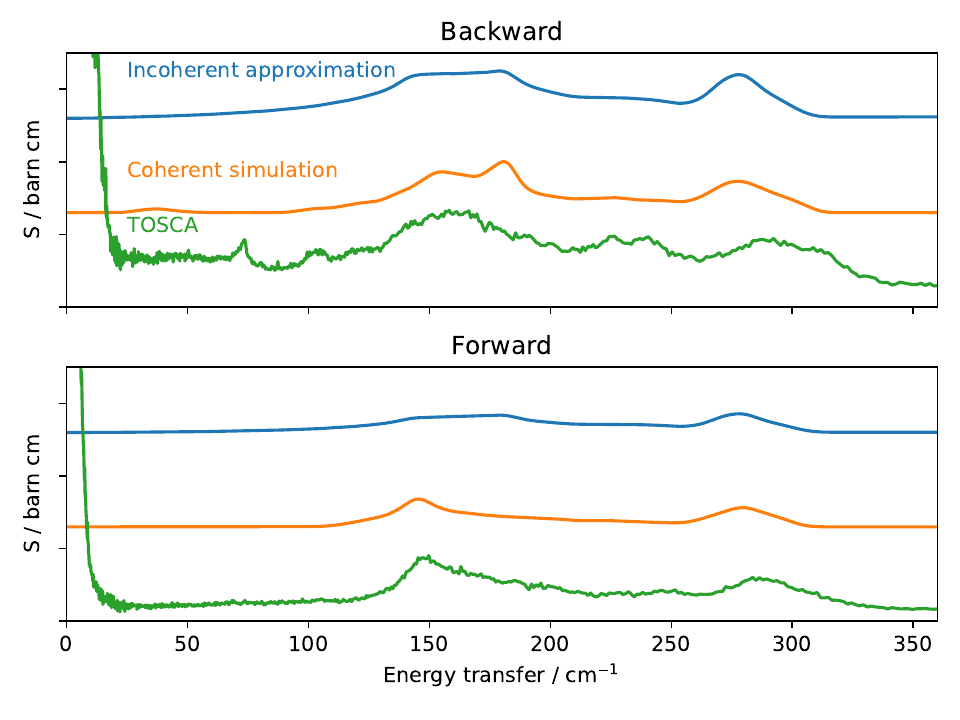}
        \label{fig:al-comparison}
    }
        
    \subfloat[\sicSixH{} (PBEsol calculations)]{
        \includegraphics[width=0.5\textwidth]{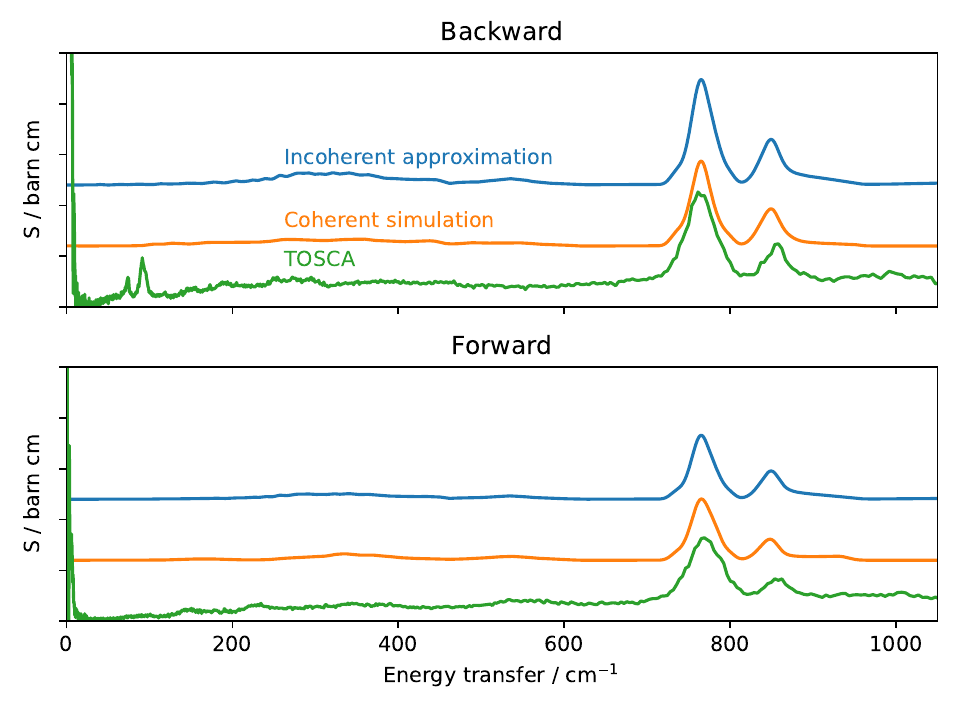}
    }

    \subfloat[\ce{Ga2O3} (RSCAN calculations)]{
        \includegraphics[width=0.5\textwidth]{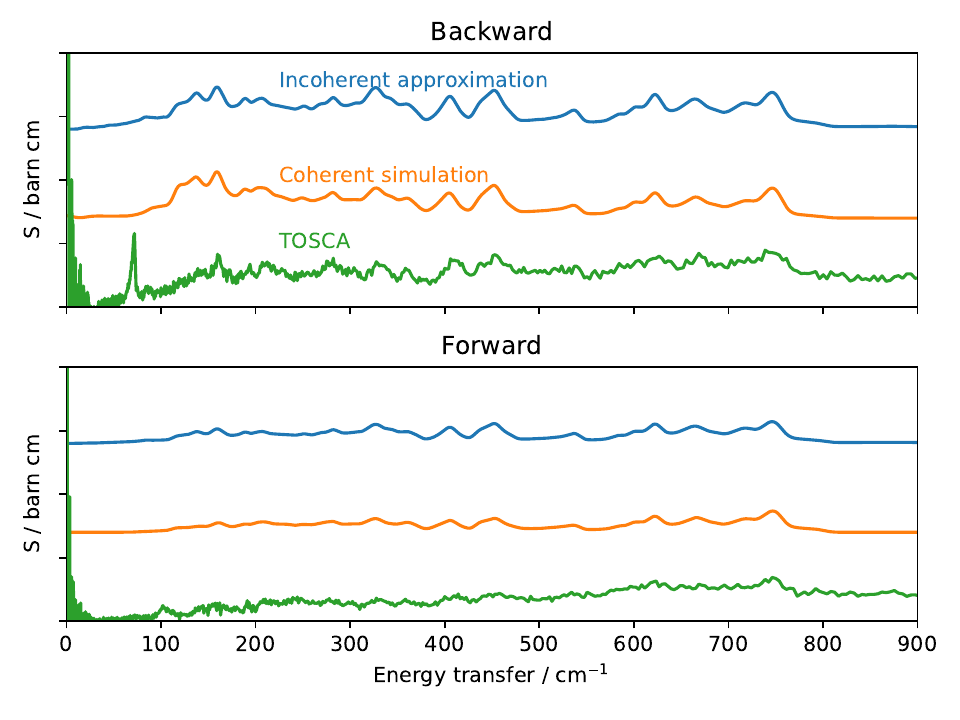}    
    }

    \caption{Comparison of low-temperature simulations in incoherent \gls{dos}-like approximation, coherent numerically-averaged calculation, and TOSCA measurements.}
    \label{fig:abins-euphonic-expt-comparisons}
\end{figure}

\abins{} simulation results are shown in \cref{fig:sic-abins,fig:ga2o3-abins}:
note that unlike \cref{fig:mari-sic-5K,fig:mari-ga2o3-5K} the dispersion information is neglected, giving rise to unphysical flat bands in the MARI simulations.
However, comparing the 1-dimensional TOSCA spectra in \cref{fig:abins-euphonic-expt-comparisons} the differences in included features are quite subtle
and we cannot confidently declare that one simulation method or the other gives a better match to the experimental data.

In the aluminium case (\cref{fig:al-comparison}) the numerically-averaged coherent simulation better captures the crisp peak close to \SI{150}{\per\cm} in forward-scattering,
but it also shows more structure in this energy region of the backscattering spectrum where the experimental data has a single broad feature.
In the \alphasic{} and \betagox{} cases the spectra are more similar and seem to capture the same features.

\section{Conclusions}
\Acrlong{ins} measurements were performed for \alphasic{}, \betagox{} and empty Al sample environments using the (indirect-geometry) TOSCA and (direct-geometry) MARI neutron spectrometers at the ISIS Neutron and Muon Source.
Phonon force constants were computed in the harmonic approximation using DF(P)T and a variety of \gls{xc} approximations, with a novel quasirandom-sampling scheme used to evaluate reciprocal-space convergence.
With these force constants \gls{ins} spectra were simulated, accounting for instrumental resolution and energy/momentum restrictions.
Two established methods were compared:
a \gls{dos}-like incoherent approximation typically used for molecular crystals on TOSCA,
and a numerically-averaged coherent sampling typically used for inorganic powders on MARI.
Novel simulations of TOSCA spectra were performed by finely slicing $(Q,\omega)$ samples from the full 2-D intensity map and applying instrumental resolution.
We find that despite the small unit cells and coherent scattering nuclei, the incoherent \gls{dos}-like approximation performs acceptably as a model of \gls{ins} on TOSCA for these materials.
In 2--D MARI spectra the phonon dispersion effects are visible and were successfully reproduced in simulation by numerical sampling of the coherent phonon scattering in constant-$Q$ shells.

For both ceramics, the frequencies from \gls{gga} and \gls{lda} are similar while the RSCAN frequencies are generally higher.
The \alphasic{} measurements are closely matched by PBEsol calculations in the 6H polytype,
which is consistent with the established reliability of PBEsol for SiC lattice parameters and bulk modulus.
For \betagox{} we find that RSCAN gives significantly better agreement with experiment than other tested functionals.
For \abinitio{} research into the phonon-mediated low thermal conductivity of \betagox{}, we recommend the use of RSCAN or similar \gls{metagga} functionals to obtain good agreement with experiment.
Given the increased cost of such calculations, \glspl{mlip} could be used as a surrogate model to extend the \gls{metagga} simulations to larger regions and longer timescales;
this approach has already been employed to reproduce existing \gls{ins} measurements of other materials~\cite{lindgrenPredictingNeutronExperiments2025}.

\gls{ins} experiments are usually performed at cryogenic temperatures to minimise the Debye--Waller intensity loss which is driven by atomic motions.
For these dense semiconductors with relatively low phonon frequencies (compared to organic stretching and bending modes) the effect of temperature was quite subtle.
This suggests that it could be productive in high-throughput study of similar materials to operate at ambient temperature and omit equilibration time.

\section*{Data Access Statement}
A full set of experimental and simulated MARI spectra are included as \suppl{} with this article \cite{supp:material}.
CASTEP force constants are available with input files from STFC eData at \url{https://doi.org/10.5286/edata/972}.
A Python/snakemake workflow is made available at \url{https://doi.org/10.5281/zenodo.20349290} which implements the described method and allows plots like \cref{fig:sliced} to be produced from force-constants data in CASTEP or Phonopy format.
The \abins{} \qpoint{} convergence plots in SI were produced with a reusable workflow available from \url{https://doi.org/10.5281/zenodo.13902655}.
Data from the ISIS experiments is available at \url{https://doi.org/10.5286/ISIS.E.RB2010450} (MARI) and \url{https://doi.org/10.5286/ISIS.E.RB2010453} (TOSCA).

\section*{Acknowledgements}
We acknowledge useful discussions with Dominik Jochym, Keith Refson, Jacob Wilkins, Richard Waite and Ciar\'{a}n O'Brien.
Experiments at the ISIS Neutron and Muon Source were supported by beamtime allocations RB2010450 and RB2010453 from the Science and Technology Facilities Council. 
Computing resources were provided by STFC Scientific Computing's  SCARF cluster and STFC Cloud service.

\bibliographystyle{unsrt}
\bibliography{bibliography}
\end{document}